\documentclass{PoS}

\title{Phase transitions in the three-dimensional $Z(N)$ models}

\ShortTitle{Phase transitions in the three-dimensional $Z(N)$ models}

\author{Oleg Borisenko	\\
        BITP, National Academy of Science of Ukraine, 003680 Kiev, Ukraine\\
        E-mail: \email{oleg@bitp.kiev.ua}}
				
\author{\speaker{Vladimir Chelnokov} \\
        BITP, National Academy of Science of Ukraine, 003680 Kiev, Ukraine\\
        E-mail: \email{vchelnokov@i.ua}}
				
\author{Gennaro Cortese	\\
	Instituto de F\'isica Te\'orica UAM/CSIC, Cantoblanco, E-28049 Madrid, Spain, \\
	\& Departamento de F\'isica Te\'orica, Universidad de Zaragoza, E-50009 Zaragoza, Spain \\
        E-mail: \email{cortese@unizar.es}}
				
\author{Mario Gravina	\\
	Dipartimento di Fisica, Universit\`a della Calabria, \\
	\& INFN - Gruppo Collegato di Cosenza, I-87036 Rende, Italy \\
				E-mail: \email{gravina@cs.infn.it}}
				
\author{Alessandro Papa	\\
	Dipartimento di Fisica, Universit\`a della Calabria, \\
	\& INFN - Gruppo Collegato di Cosenza, I-87036 Rende, Italy \\
				E-mail: \email{papa@cs.infn.it}}
								
\author{Ivan Surzhikov	\\
        BITP, National Academy of Science of Ukraine, 003680 Kiev, Ukraine\\
        E-mail: \email{i\_van\_go@inbox.ru}}
				
\abstract{Phase transitions in zero-temperature $3D$ $Z(N)$ lattice gauge theories 
are studied. We use a cluster algorithm defined for the dual formulation of the 
models. We also attempt to explain the nature of the intermediate continuously 
symmetric phase, which appears for $N>5$. 
The critical indices are calculated. The results obtained are used to study the
scaling of critical points with $N$, as well as the scaling of finite-temperature 
critical points with the lattice size in the time direction, $N_T$.}

\FullConference{31st International Symposium on Lattice Field Theory LATTICE 2013\\
		 July 29 - August 3, 2013\\
		 Mainz, Germany}

\begin{document}

\section{Introduction}
The properties of the phase transitions in $Z(N)$ models are interesting 
both by themselves and due to their connection with the deconfining transitions 
in $SU(N)$ gauge theories, $Z(N)$ being the center subgroup of $SU(N)$. 
In this paper we study the $Z(N)$ lattice gauge theory (LGT) at zero temperature. 
In the most general form, the action of the isotropic $Z(N)$ LGT can be written as 
\begin{equation} 
S_{\rm gauge}  =  \frac{1}{2} \  \sum_x \sum_{n<m} \ \sum_{k=1}^{N-1} \beta_k
\exp \left( \frac{2 \pi i k}{N} \left(s_n(x) + s_m(x+e_n) 
-s_n(x+e_m) - s_m(x) \right) \right) \ ,
\label{action_gauge}
\end{equation}
where $e_n$, $n=1,2,3$, denotes the unit vector in the $n$-th direction. 
Similarly, the most general action of the $Z(N)$ spin model is given by  
\begin{equation} 
S_{\rm spin} \ = \ \frac{1}{2} \ \sum_x \sum_{n} \ \sum_{k=1}^{N-1} \beta_k
\exp \left( \frac{2 \pi i k}{N} \left(s(x) - s(x+e_n) \right) \right) \ .
\label{action_spin}
\end{equation}
In both cases we used the convention
\begin{equation} 
\beta_k \ = \ \beta_{-k} \ = \ \beta_{k+N}  \ > \ 0 \ .
\label{beta_convention}
\end{equation}
The standard Potts model corresponds to the choice when all $\beta_k$ are 
equal. Then, the sum over $k$ reduces to a delta-function on the $Z(N)$ group. 
The conventional vector model corresponds to $\beta_k=0$ for all $k\ne 1,N-1$. 
For $N=2,3$ the Potts and vector models are equivalent. 

Two-dimensional (2D) standard and vector $Z(N)$ LGTs are exactly solvable both 
in the finite volume and in the thermodynamic limit. They exhibit no phase 
transition at any finite value of the coupling constant $\beta$. For $3D$ 
$Z(N)$ models no exact solution is known. While the phase structure of the general 
model defined by~(\ref{action_gauge}) remains unknown, it is well established 
that Potts and vector models with only $\beta_1$ non-vanishing have one 
phase transition from a confining phase to a phase with vanishing string 
tension~\cite{horn,ukawa,savit}. 

The duality relation between $3D$ $Z(N)$ gauge and spin models allows to establish 
the order of phase transition in $3D$ $Z(N)$ LGT using the results obtained 
for spin models. 
Since $Z(2)$ LGT is equivalent to the Ising model, its critical behavior is 
well known (see Refs.~\cite{caselle} and references therein). Generally, the 
$Z(N)$ global symmetry of the finite-temperature $4D$ $SU(N)$ gauge theory 
motivated thorough investigations, both analytical and numerical,  
of the $3D$ spin models, especially for $N=2,3$~\cite{gavai,fukugita} (for 
more recent studies, see~\cite{bazavov_z3} and references therein). 
The $3D$ Potts models for $N>3$ have been simulated in~\cite{bazavov} 
and studied by means of the high-temperature expansion in~\cite{janke}. 

Surprisingly, much less is known about the critical behavior of $Z(N)$ vector 
LGTs when $N>4$. They have been studied numerically in~\cite{bhanot} up 
to $N=20$ on symmetric lattices with size $L\in [4-16]$. It was confirmed that 
zero-temperature models possess a single phase transition which disappears in 
the limit $N\to\infty$. A scaling formula proposed in~\cite{bhanot} shows that 
the critical coupling diverges like $N^2$ for large $N$. 
Thus, the $U(1)$ LGT has a single confined phase in 
agreement with theoretical results~\cite{3d_u1}. 
We are not aware, however, of any detailed study of the critical behavior 
of the vector models with $N\geq 4$ in the vicinity of this single phase 
transition. Slightly more is known about the critical properties of $Z(N)$ 
vector spin models. In particular, it has been suggested that all vector spin 
models exhibit a single second order phase transition~\cite{scholten}. 
An especially detailed study was performed on the $Z(6)$ model, because the
$Z(6)$ global symmetry appears as an effective symmetry of the $Z(3)$ 
antiferromagnetic Potts model~\cite{z6_vector_mc,z6_vector_rg}. The computed 
critical indices suggest that the $Z(6)$ vector model belongs to the 
universality class of the $3D$ $XY$ model. An interesting feature of the $Z(6)$
model and, possibly, of all vector models with $N>4$, is the appearance of an
intermediate rotationally symmetric region below the critical temperature of the 
second order phase transition. The mass gap, however, was found to be rather 
small, but non-vanishing in this region~\cite{z6_vector_mc}. 
Combined with a renormalization group (RG) study, the analysis 
concluded that this intermediate region presents a crossover to a
low-temperature massive phase, where the discreteness of $Z(6)$ plays an 
essential role~\cite{z6_vector_rg}.      

The main goal of the present work is to fill the gap in our knowledge about 
the critical behavior of the $3D$ $Z(N)$ LGTs. Another motivation comes from 
our recent studies of the deconfinement transition in the $Z(N)$ vector LGT 
for $N>4$ at finite temperatures~\cite{3d_zn_strcoupl,lat_12,ZN_fin_T}. The 
major findings of these papers was the demonstration of two phase transitions 
of the Berezinskii-Kosterlitz-Thouless type and the existence of an 
intermediate massless phase. The critical indices at these transitions have 
been found to coincide with the indices of the $2D$ vector spin models. An 
interesting question then arises regarding the construction of the continuum 
limit of the finite-temperature models in the vicinity of the critical points. 
For this to accomplish it might be useful, and even necessary, to know the 
scaling of quantities such as string tension, correlation length, 
{\it etc.} near the critical points of the corresponding zero temperature 
models.  

\section{Numerical results}

To study the phase transitions in $3D$ $Z(N)$ LGTs, it turns out to be more 
convenient to simulate the dual spin model, whose action is given in~(\ref{action_spin}). The equations for obtaining the coupling constants 
for the dual spin model are given in~\cite{3dzn0}.
Simulations were performed by means of a cluster algorithm on symmetric 
lattices $L^3$ with periodic BC and $L$ in the range 8 -- 96.
For each Monte Carlo run the typical number of generated configurations 
was $2.5 \cdot 10^6$, the first $10^5$ of them being discarded to ensure 
thermalization. Measurements were taken after every 10 whole lattice updates and error 
bars were estimated by the jackknife method combined with binning.
The following observables were used:
\begin{itemize}
\item complex magnetization $M_L = |M_L| e^{i \psi}$, $M_L = \sum_{x \in \Lambda} \exp \left( \frac{2 \pi i}{N} s(x) \right) \;;$
\item population $S_L$, $S_L = \frac{N}{N - 1} \left(\frac{\max_{i = 0, N - 1} n_i} {L^3} - \frac{1}{N} \right)\;,$
where $n_i$ is number of $s(x)$ equal to $i$;
\item susceptibilities of $M_L$ and $S_L$:  
$\chi_L^{(M)}$, $\chi_L^{(S)}$, $\chi_L^{(\mathbf\cdot)} \ =\  L^2  \left(\left< \mathbf\cdot^2 \right> 
- \left< \mathbf\cdot \right>^2 \right)\;;$
\item Binder cumulant $U_L^{(M)} = 1 - \frac{\left\langle \left| M_L \right| ^ 4 
\right\rangle}{3 \left\langle \left| M_L \right| ^ 2 \right\rangle^2}\;.$
\end{itemize} 
We computed also the heat capacity in the vicinity of 
the critical points. 

To obtain the critical couplings, the Binder cumulant crossing method
described in~\cite{3dxy_univ} was used. In particular, we computed by Monte Carlo
simulations the Binder cumulant $U_L^{(M)}$ and its first three derivatives
with respect to $\beta$ for the different lattice sizes, thus allowing 
to build the function $U_L^{(M)}(\beta)$ in the region near the transition. Then, 
we looked for the value of $\beta$ at which the curves $U_L^{(M)}(\beta)$
related to the different lattice sizes $L$ ``intersect''. In fact, 
the critical coupling $\beta_{\rm c}$ was estimated as the value of 
$\beta$ at which $U_L^{(M)}(\beta)$ exhibits the least dispersion over 
lattice sizes ranging from $L=16$ to $L=96$. To check critical couplings we have compared them 
with the results of the SDA RG~\cite{sdarg} and found out that the values obtained from RG and from Monte Carlo simulations are very close.

The critical coupling values for the $3D$ $Z(N>4)$ vector models can be fitted with the formula 
\begin{equation}
\beta_{\rm c} = \frac{A}{1-\cos\left(\frac{2 \pi}{N}\right)} 
+ C \left({1-\cos\left(\frac{2 \pi}{N}\right)}\right) \ ,
\label{fit_betac}
\end{equation}
being a generalization of the critical coupling scaling given in~\cite{bhanot}, with following parameters
$A=1.50122(7)$, $C=0.0096(5)$, $\chi^2/{\rm d.o.f.} = 13.1$. Despite the large $\chi^2$, 
probably due to the underestimation of the error bars of critical
couplings, the proposed function nicely interpolates data over a large
interval of values of $N$. 

The procedure to determine the critical index $\nu$ is also inspired 
by Ref.~\cite{3dxy_univ}: for each lattice size $L$ the known function
$U_L^{(M)}(\beta)$ is used to determine $d U_L^{(M)}(\beta)/d\beta$; from this, the
derivative of $U_L^{(M)}$ with respect to the rescaled coupling 
$x = (\beta - \beta_{\rm c}) L^{1/\nu}$ can be calculated,
\begin{equation}
\frac{d U_L^{(M)}} {d x} = \frac{d U_L^{(M)}}{d \beta} L^{1/\nu}\;.
\label{nu_determ}
\end{equation}
The best estimate of $\nu$ is found by minimizing the deviation of 
$d U_L^{(M)}/d x$ with respect to a constant value. The minimization can be done
at $\beta_{\rm c}$ or at any other value $\beta_{\rm f} \approx \beta_{\rm c}$ 
defined as the point where $U_L^{(M)}$ on a given lattice becomes equal to some 
fixed value. The resulting values for $\nu$, do not differ within error bars.

\begin{figure}
\centering
\includegraphics[width=0.49\textwidth]{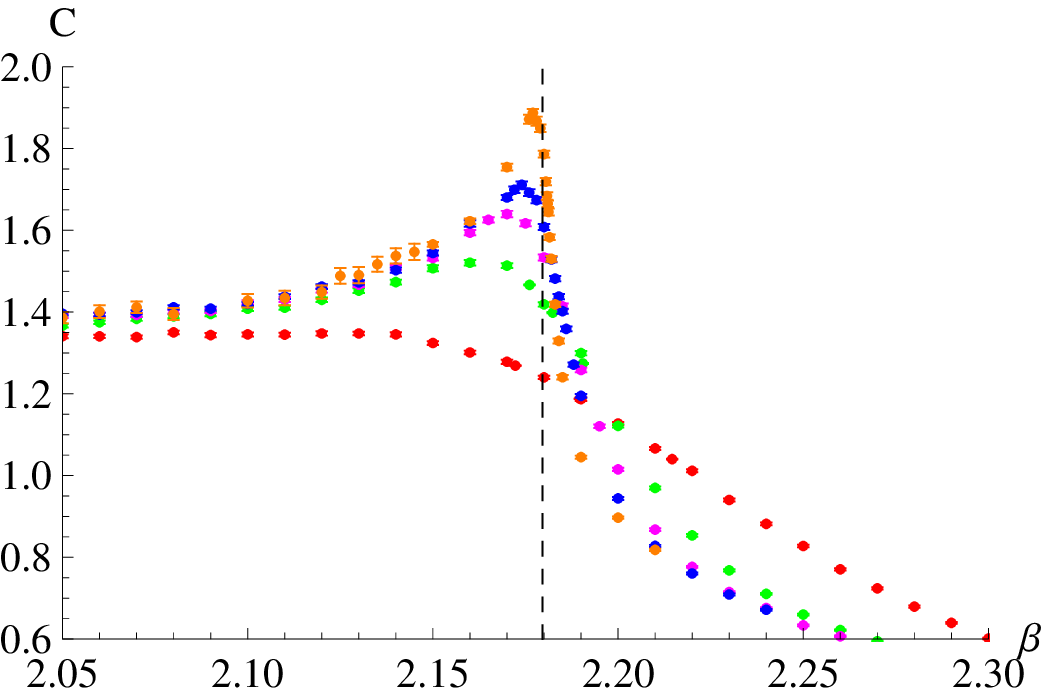}
\includegraphics[width=0.49\textwidth]{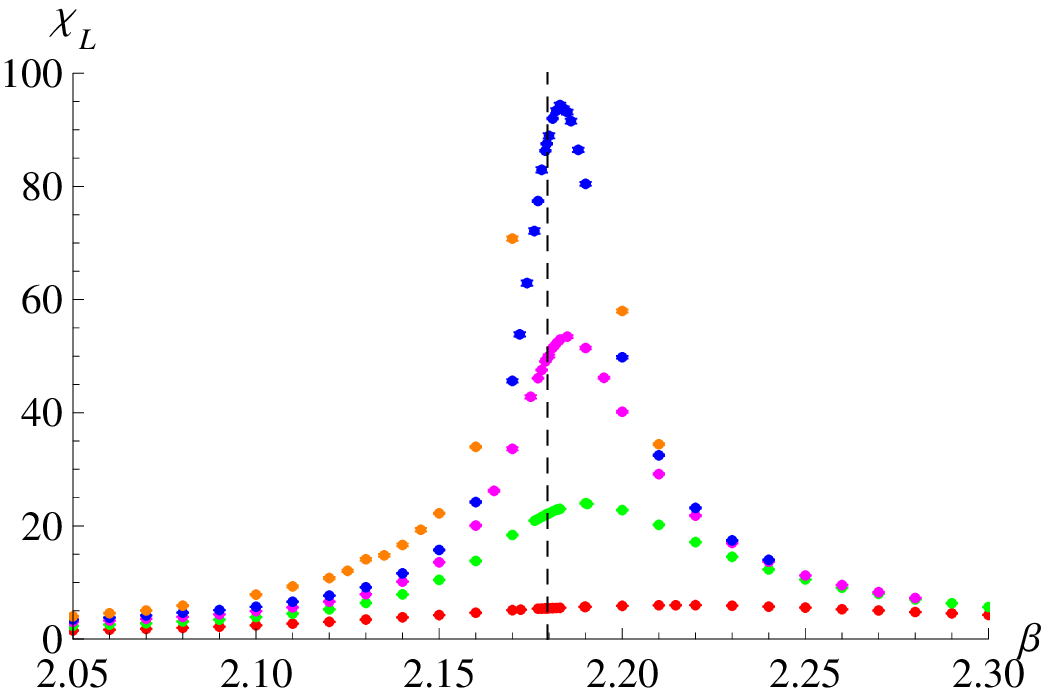}
\caption{Heat capacity (left) and susceptibility $\chi_L^{(M)}$ (right) 
{\it versus} $\beta$ for the $3D$ $Z(5)$ vector model on lattices with 
$L$ = 8 (red), 16 (green), 24 (violet), 32 (blue) and 64 (orange). The 
vertical dashed line shows the location of the critical point. The error bars
are smaller than symbols.}
\label{spchiplots}
\end{figure}

The critical index $\alpha$, determined from the $\nu$ values obtained
in the previous subsection by means of the relation $\alpha=2-d\nu$, gets
negative values for all $N\geq 5$, meaning that the transition is of order higher than two. 
In fact, these negative values are very close to that of the $3D$ $XY$ model~\cite{3dxy_univ}.  
However, the plots of the heat capacity (see Fig.~\ref{spchiplots}(left)) 
clearly show that it diverges in the vicinity of the critical point. 
Moreover, the maxima of the heat capacity and of the susceptibility $\chi_L^{(M)}$ 
approach the critical point from different sides (see Figs.~\ref{spchiplots}).

This suggests that a different value for the index $\nu$ can be found
if a finite-size scaling (FSS) analysis is done on the peak values of the heat 
capacity, using as fitting function
\begin{equation}
C(L) = A L^{\alpha/\nu}\ .
\label{hc_fit}
\end{equation} 
After $\alpha/\nu$ is extracted, from the relation $\alpha=2-d \nu$
the value of $\nu$ can be obtained.

We see that while for $N=2, 4$ the resulting $\nu$ agrees with the value of 
$\nu$ in the $3D$ Ising model, it is not so for $N>4$. The most important
fact is, however, that in all cases the $\nu$ indices obtained this way
are close to $\nu \approx 0.63$ -- the critical index for the Ising model. 
The difference between $\nu$ indices obtained from the $U_L^{(M)}$ cumulants 
and from the heat capacity leads us to conclude that we have two kinds of 
singularity depending on whether one approaches the critical coupling from 
above ($3D$ $XY$ model-like singularity) or from below 
($3D$ Ising universality class), for $N > 4$. 

Finally, the critical indices $\beta/\nu$ and $\gamma/\nu$ can be extracted from 
the FSS analysis of the magnetization $M_L$ and its
susceptibility $\chi_L^{(M)}$, according to the following fitting functions,
\begin{equation}
\label{fit_def}
M_L = A_1 L^{-\beta/\nu} \ , \ \ \chi_{M_L} = A_2 L^{\gamma/\nu}(1 + B_2 L^{-\delta/\nu}) \ .
\end{equation}
The critical index $\eta$ will then be given by $2-\gamma/\nu$ and the hyperscaling relation
$d=2\beta/\nu+\gamma/\nu$ must be satisfied with $d=3$.
The $\delta$ exponent was fixed to be 0.53 -- the value for the $XY$ model. When we take $\delta$ in the range 0.4-0.7 the change of the critical index $\gamma/\nu$ 
remains smaller than the error estimates, so our results do not depend much on the exact value of $\delta$. On the other hand, without this second order correction
$\eta$ value becomes much smaller (see~\cite{3dzn0} for comparison). 

The critical indices for different $Z(N)$, $N = 2,4,5,8,13,20$ are summarized in Table~\ref{results1}. More
detailed results with comparison of the different methods can be found in \cite{3dzn0}. 

\begin{table}[ht]
\caption{Critical couplings and critical indices for $3D$ $Z(N)$ models: ($\nu_{\rm Binder}$ corresponds to 
the $\nu$ index obtained from the Binder cumulant derivatives, $\nu_{\rm heat}$ -- the $\nu$ index obtained from the heat capacity, $d = 2\beta/\nu + \gamma/\nu$ and 
should be equal to 3).
}
\begin{center}
\begin{tabular}{|c|c|c|c|c|c|c|}
\hline
$N$ & $\beta_{\rm c}$ & $\nu_{\rm Binder}$ & $\nu_{\rm heat}$ & $\beta/\nu$ & $\gamma/\nu$ &  $\eta$ \\
\hline
2  & 0.761395(4) & 0.6306(8)   & 0.6143(6)  & 0.504(2)  & 1.975(8)  & 0.025(18) \\
4  & 1.52276(4)  & 0.62933(12) & 0.6168(14) & 0.493(2)  & 1.951(14) & 0.049(14) \\
5  & 2.17961(10) & 0.6681(8)   & 0.6360(6)  & 0.5088(7) & 1.971(12) & 0.029(12) \\
6  & 3.00683(7)  & 0.6756(16)  & 0.6360(10) & 0.5052(9) & 1.970(5)  & 0.030(5) \\
8  & 5.12829(13) & 0.6748(2)   & 0.6336(3)  & 0.5083(8) & 1.966(4)  & 0.034(4) \\
13 & 13.1077(3)  & 0.6723(17)  & 0.6340(4)  & 0.5092(11)& 1.971(7)  & 0.029(7) \\
20 & 30.6729(5)  & 0.6739(7)   & 0.6314(3)  & 0.5034(7) & 1.99(6)   & 0.01(6)  \\
\hline
\end{tabular}     
\end{center}
\label{results1}
\end{table} 

\section{Symmetric phase}

Another interesting phenomenon we have encountered during our study is the 
appearance of a symmetric phase just below $\beta_{\rm c}$ for all $N\geq 5$. 
That such phase exists in the vector $Z(6)$ spin model has been known for a 
long time~\cite{scholten,z6_vector_mc,z6_vector_rg}. 
Here we confirm its existence for all vector $Z(N\geq 5)$ LGTs. The phase 
exhibits itself, {\it e.g.}, in the distribution of magnetization angle. 
In the region just below critical point the magnetization angle is evenly distributed,
showing that the continuous symmetry is not broken. Only at much smaller couplings 
one can observe the appearance of a symmetry-broken phase. 
In addition, we have studied the behavior of the population susceptibility 
below $\beta_{\rm c}$. For $N \geq 5$ it has a second 
broad maximum, which slowly moves to $\beta_{\rm c}$ with increasing
lattice size. While, for $N=5$ the peak of the population susceptibility moves 
rather fast and practically collapses with the peak at the critical coupling 
on the largest available lattice $L=96$, for larger $N$ the peak stays 
rather far from the corresponding critical coupling, even for $L=96$ 
(with our data we cannot even exclude a situation when the convergence of 
the second maximum is logarithmic). 
We can imagine two scenarios to explain such behavior: 
\begin{enumerate} 
\item This symmetric phase exists only in finite volume. When the lattice size 
increases, the second maximum approaches the critical coupling and, eventually,
the symmetric region shrinks and disappears. 
The explanation proposed in~\cite{z6_vector_mc,z6_vector_rg} might work in 
this case, too. 
Namely, the symmetric phase on the finite lattice is a phase with a very small 
mass gap and describes a crossover region to the symmetry-broken phase. 
\item For $N>5$ the second maximum of the population susceptibility stays 
away from the critical couplings even in the infinite volume limit. In this 
case it might correspond to some higher-order phase transition 
and the symmetric phase with tiny or even vanishing mass gap exists also in 
the thermodynamic limit.
\end{enumerate}
In both cases it is tempting to speculate that this symmetric region is 
reminiscent of the massless phase which appears in these models at finite 
temperature~\cite{ZN_fin_T,3dznfin}. 
Whichever scenario of the above two is realized, one needs to study the 
models on much larger lattices to uncover it. 

\section{Summary} 

We have studied the $3D$ $Z(N)$ LGT at zero temperature 
aiming at shedding light on the nature of phase transitions in these models 
for $N\geq 4$. This study was based on the exact duality transformations of 
the gauge models to generalized $3D$ $Z(N)$ spin models.

The main results can be shortly summarized as follows: 
\begin{itemize}
\item We have determined numerically the position of the critical couplings 
for various $Z(N)$ models. 
For $N=2,3$ we find a reasonable agreement with the values quoted in the 
literature. For larger $N$, we have significantly improved the values given 
in~\cite{bhanot}. Critical points obtained are in good correspondence with 
the results of our RG estimations~\cite{sdarg}.

\item The critical indices $\nu$ and $\eta$ derived here for $N=2,4$ suggest 
that these models are in the universality class of the $3D$ Ising model, 
while our results for all $N>4$ hint all vector $Z(N\geq 5)$ LGTs 
belong to the universality class of the $3D$ $XY$ model. The value of the index $\nu$
 stays very close to the  $XY$ value, $\nu\approx 0.6716$, given in~\cite{3dxy_univ}. 
The index $\alpha$ in this case takes a small negative value. It thus follows 
that a third order phase transition takes place for $N\geq 5$.    

\item The index $\alpha$ extracted from the specific heat roughly agrees with 
the value of $3D$ Ising model for all $N$ studied. The fact that we observe 
two different values of the index $\alpha$ dependently on whether we approach 
the critical point from below or from above leads to the conclusion 
that the first derivative of the free energy could exhibit a cusp in 
the thermodynamic limit if $N>4$. 

\item Our data also revealed the existence of a symmetric phase for all $Z(N)$ 
vector LGTs if $N>4$. However, substantially larger lattices are required 
to see if this phase survives the transition to the thermodynamic limit. 

\item The scaling $a T_c = (\beta_{\rm c} - \beta_{\rm c}(N_T))^{-\nu} / N_T$ can be used 
to describe the dependence of finite-temperature critical points obtained in~\cite{3dznfin}
on $N_T$.

\end{itemize}

\end{document}